\documentstyle[12pt,aasms4]{article}
\def\lesssim{\mathrel{\hbox{\rlap{\hbox{\lower4pt\hbox{$\sim$}}}\hbox{$<$}}}}
\def\gtrsim{\mathrel{\hbox{\rlap{\hbox{\lower4pt\hbox{$\sim$}}}\hbox{$>$}}}}

\begin{document}

%\thesaurus{1(1;2;3)}
\title{Accretion Disk Models with Self-Consistent Advective Cooling and Heating}
 \author{I.V.Artemova\altaffilmark{1},
 \affil{Theoretical Astrophysics Center, Blegdamsvej 17, DK-2100, Copenhagen
\O, Denmark}
 %\and
 G.S. Bisnovatyi-Kogan\altaffilmark{2},
 \affil{Space Research Institute, Profsoyuznaya 84/32, 117810 Moscow, Russia}
 %\and
 G. Bj\"{o}rnsson\altaffilmark{3},
 \affil{Science Institute, Dunhagi 3, University of Iceland, IS-107 Reykjavik,
Iceland}
 %\and
 I.D. Novikov\altaffilmark{4} }
 \affil{Theoretical Astrophysics Center, Blegdamsvej 17, DK-2100, Copenhagen
\O, Denmark}
 \affil{University Observatory, {\O}ster Voldgade 3, Dk-1350, Copenhagen \O,
Denmark}
 \affil{NORDITA, Blegdamsvej 17, DK-2100 Copenhagen \O, Denmark}
 \affil{Astro Space Center of P.N. Lebedev Physical Institute, Profsoyuznaya
84/32,
 	117810 Moscow, Russia}
 \altaffiltext{1}{e-mail: julia@nordita.dk}
 \altaffiltext{2}{e-mail: Gkogan@ESOC1}
 \altaffiltext{3}{e-mail:gulli@raunvis.hi.is}
 \altaffiltext{4}{e-mail:novikov@nordita.dk}

\begin{abstract}
We present solutions to the accretion disk structure  equations in which advective 
cooling is accounted for in a self-consistent way.
 
It is shown that for high rates of accretion, above a critical value, when without 
advection there are no solutions extending continuously from large to small radii, there exist global solutions 
with {\em self-consistent} advective cooling. These solutions are quite different from
those solutions obtained earlier by other authors, where advective cooling was not 
treated self-consistently.
\end{abstract}
 

\section{Introduction}
\noindent
Since the end of the seventies it has been realized that for high rates of accretion, $\dot{M}$,
the advection of energy with the accretion flow can play a crucial role in the inner parts of an 
accretion disk around black holes. 

In the last few years a set of papers were published, where it was shown that neglecting the advective 
heat transport at high $\dot{M}$ leads to qualitatively wrong conclusions about the topology of the 
family of solutions to the system of disk structure equations (see for example Abramowicz et al
1995; Chen et al 1995; Artemova et al 1996).

The disk structure equations without advection give rise two branches of solutions: optically thick 
and optically thin, which do not intersect for 
$L<L_{b} \leq (0.6-0.9) L_{\rm Edd}$ for $\alpha=1$ 
%$L<L_{b} \approx 0.75 L_{\rm Edd}$ for $\alpha=1$ 
and $M_{BH}=10^{8}M_{\odot}$, where $L_{\rm Edd}$ is the Eddington luminosity (Artemova et al 1996).
For larger luminosities there are no solutions of these equations extending continuously from large to small radii.

It was argued by Artemova et al (1996), that for luminosities larger than $L_b$, advection 
becomes important and induces a transition from the optically thick to the optically 
thin disk in the global solution. Thus, if we were to take advection into account, solutions to the 
disk structure equations should also exist at $L>L_{b}$. In that case the disk should consist of an
optically thin inner part, smoothly connected to an optically thick outer part.

The goal of this {\em letter} is to construct explicitly accretion disk models taking advective 
heat transport into account {\em self-consistently} and to demonstrate in such a way the correctness of 
our hypothesis, as proposed in the paper by Artemova et al (1996).

We use from now on geometric units with $G=1$, $c=1$, and use $r$ as the 
radial coordinate scaled to $r_g =GM/c^{2}$.

\section{The Model and the Method of Solution}

\noindent
In describing accretion disks when advective cooling is important, one can use the following model: 
We assume that advective cooling is added in a {\em self-consistent} way to other (local) cooling 
mechanisms in a Keplerian, geometrically thin disk. We neglect the radial momentum equation
and suggest that the disk thickness is everywhere small. This simplification is not too serious
because neglecting these factors leads to quantitative corrections only, while neglecting advection
cooling leads to qualitatively different solutions. This was discussed in detail by Abramowicz et al (1995)
and Chen et al (1995). Their treatment of advective cooling was not {\em self-consistent}, however, as their
advection parameter, $\xi$, was assumed to be constant throughout the disk as well as being independent of 
the accretion rate.

Thus we assume that the disk is geometrically thin and the rotational velocity is Keplerian under
the pseudo-Newtonian potential proposed by Paczy\'nski and Wiita (1980), $\Phi=-M/(r-2)$.
We normalize the accretion rate as $\dot{m}=\dot M/\dot M_{\rm Edd}$, where
$\dot M_{\rm Edd}=L_{\rm Edd}/c^2=4\pi M m_p/\sigma_T$.

The non-relativistic disk structure equations without advection, in the case
of the Newtonian potential,
together with the method of their solutions have been described earlier (see Artemova et al 1996).  
Here we will outline the main difference related to the use of the potential of Paczy\'nski and Wiita 
and to the inclusion of the advective cooling process in the energy balance.  
Paczy\'nski and Bisnovatyi-Kogan (1981) were the first to describe a method to account for the 
advective cooling in the disk accretion problem.

The angular velocity for the Paczy\'nski and Wiita potential is $\Omega=[M/r(r-2)^{2}]^{1/2}$.  
Recall that conservation of angular momentum for a steady-state accretion in the $\alpha$-disk 
model can be written as

\begin{equation} 
\dot{m} \left(r_g\Omega\right) \frac{3}{2}\left|\frac{d\ln\Omega}{d\ln r}\right|^{-1}f 
= \left(\frac{\sigma_{\rm T}}{m_{\rm p}}\right)h \alpha P,
\end{equation}
where the factor, $f=(1-r_{\rm in}^2\Omega(r_{\rm in})/(r^2\Omega(r)))$,
accounts for the boundary condition at the inner edge of the disk, $r_{in}$.
Neglecting advection we have $r_{in}=6$. The half thickness of the disk is denoted by $h(r)$, 
and $P(r)$ is the total pressure in the equatorial plane of the disk. 

The viscous heating rate per unit area, $Q_{+}$, is given by the formula (see e.g. 
Bisnovatyi-Kogan, 1989; Frank, King \& Raine, 1992)
\begin{equation}
\left(\frac{r_g\sigma_{\rm T}}{3 m_{\rm p}}\right)Q_{+}=\dot m\left(r_g\Omega\right)^2 
 \frac{2}{3}\left|\frac{d\ln\Omega}{d\ln r}\right| f.
\label{eq:heat}
\end{equation}

The advection cooling rate can be written in the form (see Chen and Taam 1993):
\begin{equation}
Q_{\rm adv}=-\frac{\dot M}{2\pi r} T \frac{dS}{dr}
=\frac{\dot M}{2 \pi r^2} \frac{P}{\rho} \xi,
\end{equation}
where $T(r)$ is the temperature, $S(r)$ is the specific entropy, and $\xi(r)$ is a factor which 
characterizes the advective cooling. This factor can be written in the form:
\begin{equation} 
\xi\equiv -\frac{r\rho}{P}T\frac{dS}{dr}=-\frac{r\rho}{P}\left[\frac{dE}{dr} + P\frac{dv}{dr}\right],
\label{eq:g3}
\end{equation}
where, $E$ is the energy per unit mass of the gas, $\rho$ is the matter
density and $v=1/\rho$.
The energy equation has the form $Q_{+}=Q_{\rm adv}+Q_{\rm loc}$, where $Q_{\rm loc}$ is the rate 
of all local cooling processes (see Artemova et al 1996).

All other equations are the same as in Artemova et al (1996). The principal
difference here is that now the energy balance is described by a 
differential equation rather than an algebraic one.

The boundary conditions for this system are the following: at large, $r>>100r_{g}$, the solution must coincide 
with the solution obtained without advection, and at the inner disk boundary
we should not require that $r_{\rm in}=6$,
but rather we treat $r_{\rm in}$ as an eigenvalue of the problem and allow it to be less than $6$.

We solved this system of equations by the method of subsequent iterations, with fixed $\dot{m}$ and $\alpha$, 
and varying $r_{\rm in}$ to obtain a {\em self-consistent} solution. We did this in two steps: started one calculation 
from large radii and another at the inner edge of the disk and adjusting $r_{\rm in}$ until the solutions matched 
near $r\approx 13$.

\section{Results and Discussion}

\noindent
We will compare the solutions with and without advection cooling.

In Figure 1 we plot the optical depth $\tau_{0}$ ($\propto \Sigma$, the disk surface density) as a function
of radius, $r$, for the case $M_{BH}=10^{8}M_{\odot}$, $\alpha=1$; for various values of $\dot{m}$, without
taking into account the advection cooling, i.e. we set $\xi=0$.  For accretion rates $\dot{m}<\dot{m}_{b}=14.315$,
(corresponding to $L\approx0.9L_{\rm Edd}$),
there are two families of solutions which are continuous from large radii to the inner disk edge.
For $\dot{m}>14.315$ there are no solutions for a range of radii around $r \approx 13$, thus for such high accretion 
rates there are no solutions extending continuously from large to small radii. (See also the discussion
in the paper Artemova et al, 1966).

Let us now include advection cooling. For $\dot{m}<13$ this gives rather small corrections to the solutions without 
advection. When $\dot{m}>13$ advection is essential.

In Figure 2 we compare the solutions for $\dot{m}=14.0$ and $\dot{m}=15.0$ without advection, ($\xi=0$), and with 
{\rm self-consistent} advection, $\xi=\xi(r,\dot{m})$. The functions $\xi(r,\dot{m})$ for the cases $\dot{m}_{1}=14.0$ 
and $\dot{m}_{2}=15.0$ are shown in Figure 3.  The eigenvalues of $r_{\rm in}$
are correspondingly: $r_{\rm in,\dot{m}_{1}} 
\approx 6.0$ ; $r_{\rm in,\dot{m}_{2}} \approx5.4$.

In Figure 4 we plot the distribution of pressure $P=P(r)$ for $\dot{m}_{1}=14.0$ and $\dot{m}_{2}=15.0$  In the case 
$\dot{m}_{1}=14.0$ the solution changes quantitatively, but not qualitatively. The solution is optically thick 
everywhere in the disk.

For the value $\dot{m}=15.0>\dot{m}_{b}=14.315$ there was no solution
extending continuously from large to small radii without advection, i.e. when $\xi=0$ 
(see Fig.1). With the {\em self-consistent} advection term $\xi=\xi(r)$ (see
Fig. 3) there is a solution which 
consists of an optically thin inner part that continuously turns into an
optically thick outer part (see Fig. 2).
{\em Self-consistent} solutions in this case of course also exist for higher values of $\dot{m}$.

Thus, the results presented here demonstrate that the hypothesis proposed in
the paper by Artemova et al (1996)
is correct, namely that for $\dot{m}>\dot{m}_{b}$ there are solutions to the disk structure equations with 
{\rm self-consistent} advective cooling and with transitions from optically
thick to optically thin regions. Schemes showing the topology of the families of disk solutions without advection or with 
{\em self-inconsistent} advection ($\xi=const$, for example) should be revised. The advection term, $\xi(r, \dot{m})$, 
crucially depends on $r$ and $\dot m$ (recall  Fig. 3). The opposite conclusion that $\xi$ does not depend very much 
on $\dot{m}$ was obtained in the paper by Chen (1995).

We note also that there are solutions for $\dot{m}>\dot{m}_{b}$ with {\em
self-consistent} $\xi(r)$, with optically 
thick inner part and an optically thin outer region of the disk.  We will
address these solutions as well as
the more general solution, including the radial pressure gradient and
angular velocities of the gas, in future work.

\acknowledgements

This paper was supported in part by the Danish Natural Science Research
 Council through grant 11-9640-1, in part by Danmarks
Grundforskningsfond through its support for the establishment of the
 Theoretical Astrophysical Center. G. B. thanks NORDITA and TAC, and G.B.-K. thanks TAC for
partial support and hospitality during the initial stages of this work.

\pagebreak

\clearpage
\newpage
\pagebreak
\begin{center}
FIGURE CAPTIONS
\end{center}
Fig. 1.--- {Optical depth $\tau_{0}, \;(\propto \Sigma_0)$ as a function of radius, $r$, for
$M_{BH}=10^8M_{\odot}$, $\alpha=1.0$, without advective cooling ($\xi=0$) and different values of  
$\dot{m}$. The solid curves correspond to optically thin (lower branch) and optically thick (upper) 
solutions for $\dot{m}=14.0$. The dashed curves are calculated for $\dot{m}=15.0>\dot m_b$. No solution extending 
continuously from the outer disk to its innermost edge exists in the latter case.}

Fig. 2.---  {Comparing the radial dependence of the optical depth $\tau_{0}, \;(\propto \Sigma_0)$ for models
with the same parameters as in Fig. 1. We show the solutions without advective cooling ($\xi=0$) and with 
self-consistent advective cooling. The corresponding $\xi=\xi(r, \dot{m})$
are shown in Fig. 3.
The solid curves corresponds to $\dot{m}=14.0$, $\xi=0$, the dotted curve corresponds to $\dot{m}=14.0$, 
$\xi=\xi(r,\dot{m})$, the dashed curves correspond to $\dot{m}=15.0$, $\xi=0$. 
The dot-dashed and dashed-triple dotted curve correspond to $\dot{m}=15.0$, $\xi=\xi(r,\dot{m}=15.0)$.}

Fig. 3.--- {The quantity $\xi$ as a function  of radius, $r$, and the accretion rate $\dot{m}$, for the same disk 
parameters as in Fig. 1. The dotted curve corresponds to $\dot{m}=14.0$, the dot-dashed and dashed-triple dotted 
curves correspond to $\dot{m}=15.0$.} 

Fig. 4.--- {The pressure structure of the same disk models as in Fig. 2, (in units ${\cal P}=P/\rho c^{2}$). 
The styles of the curves are the same as in Fig. 2.}

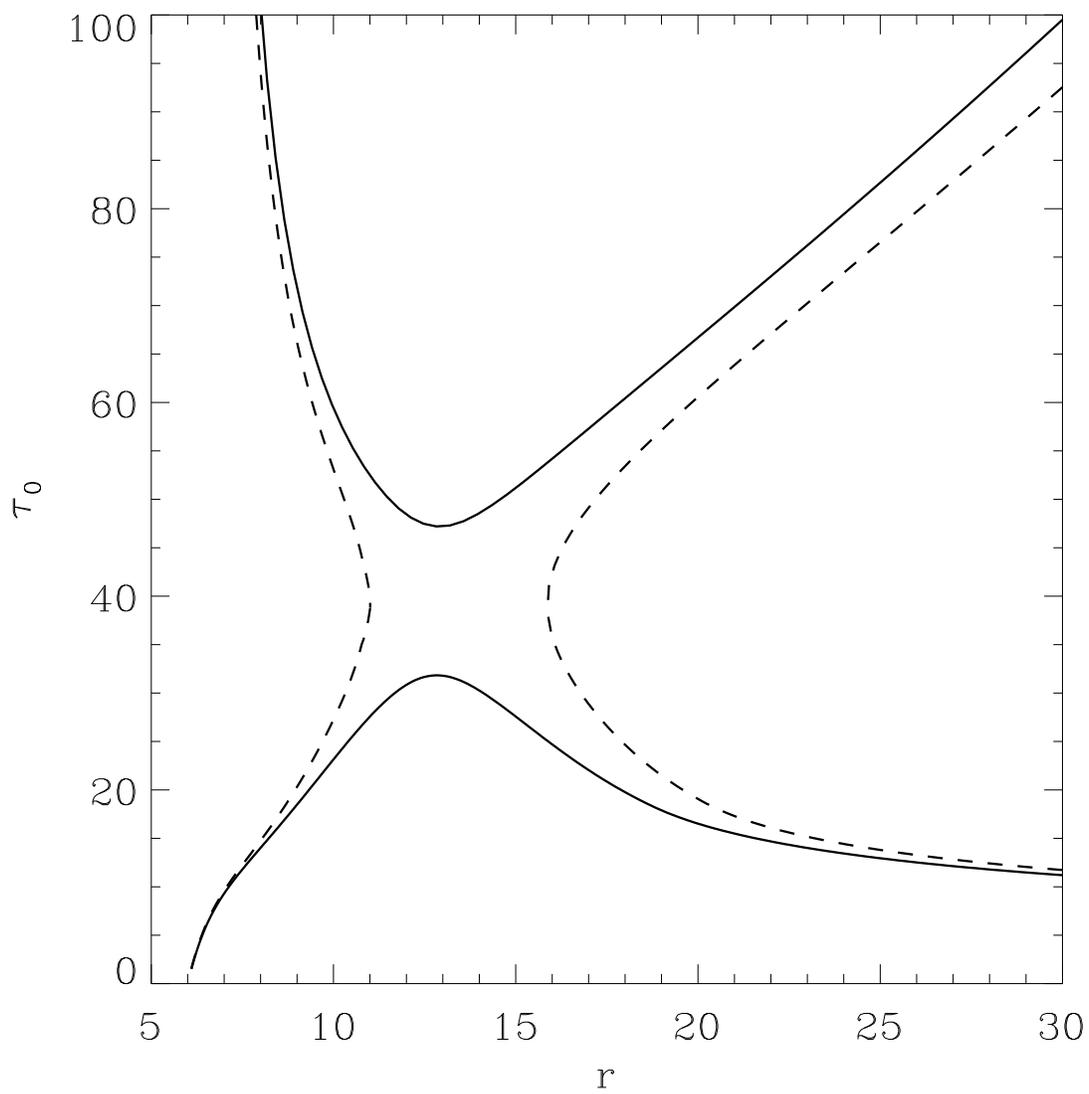
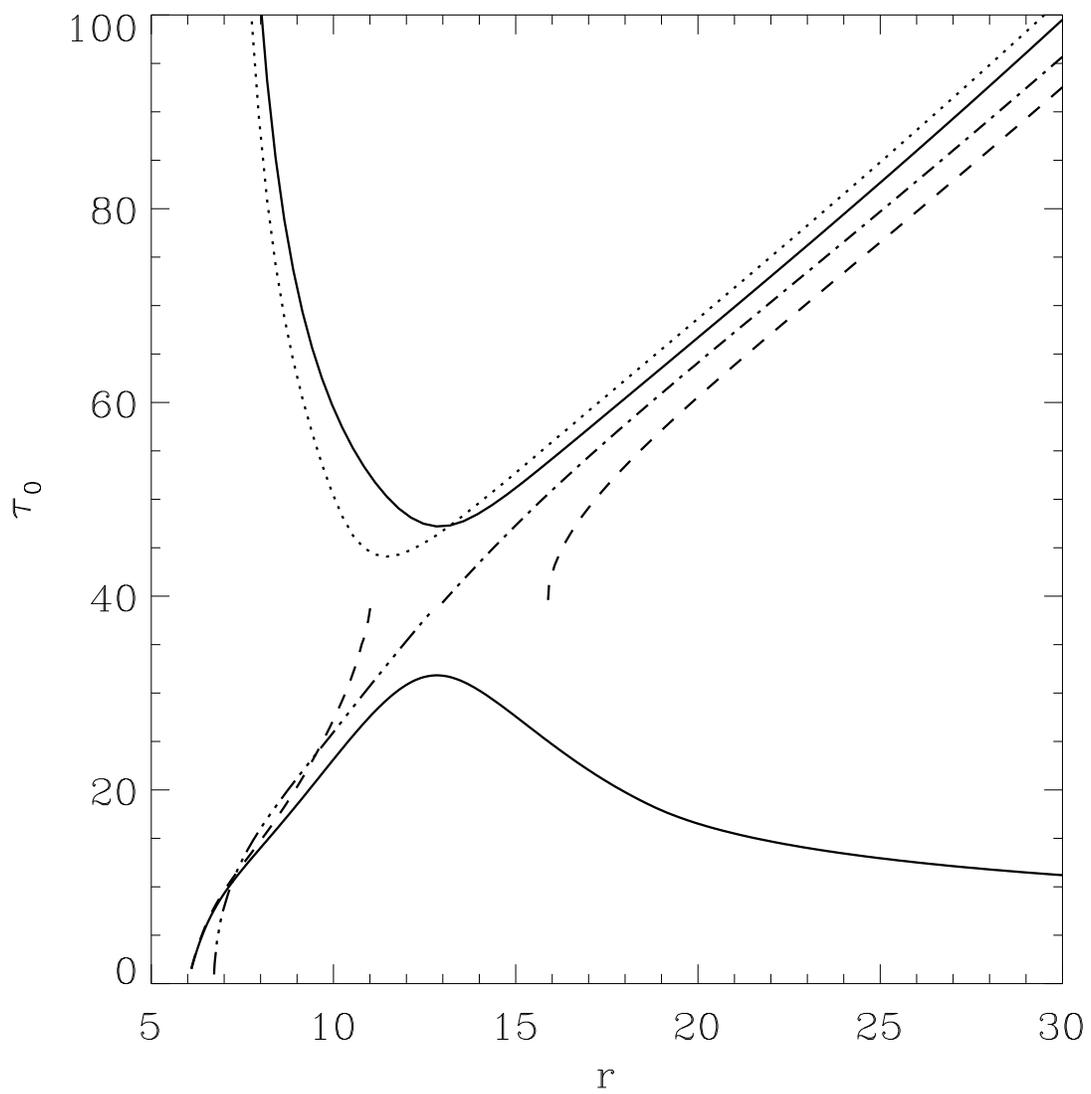
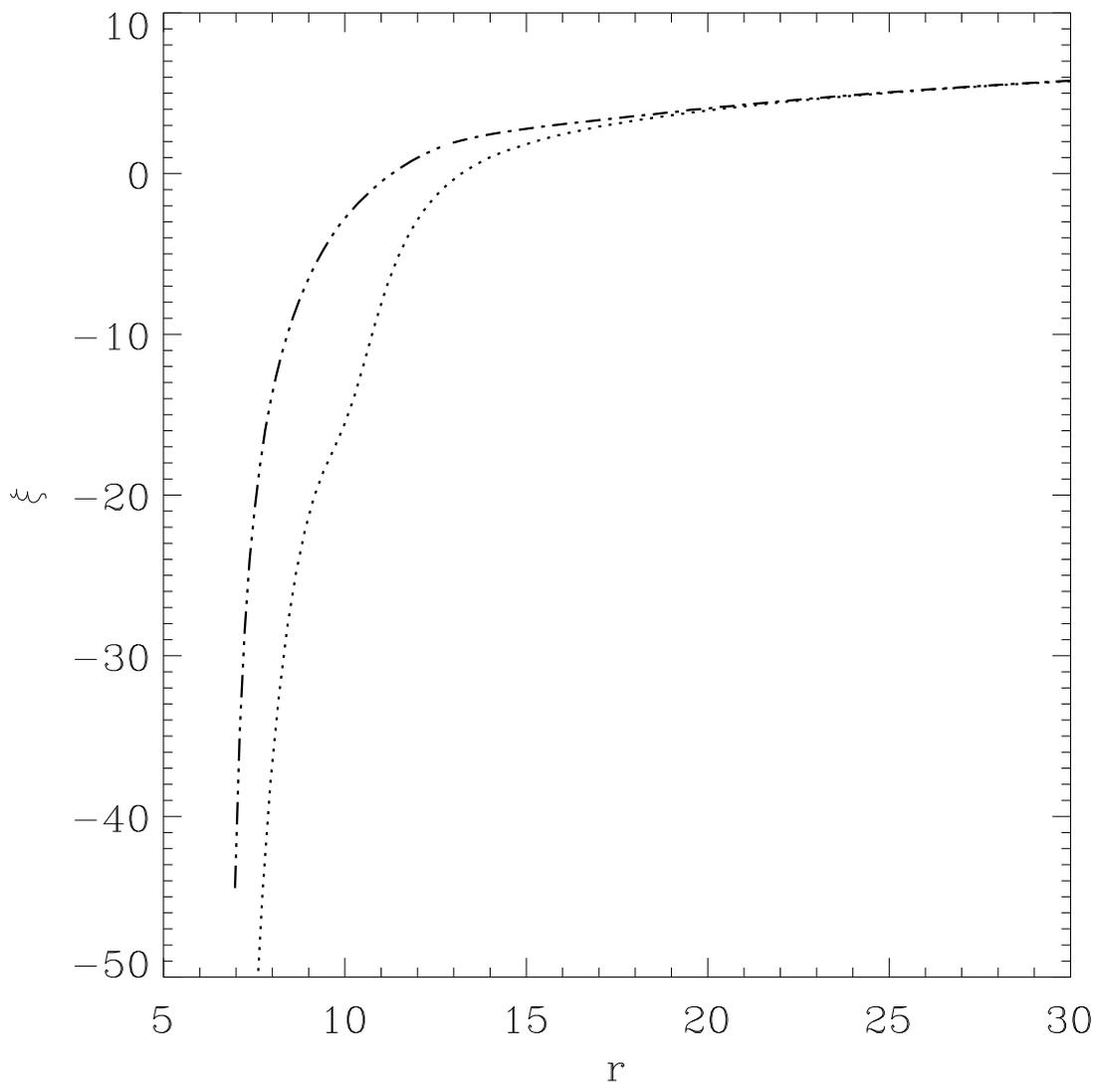
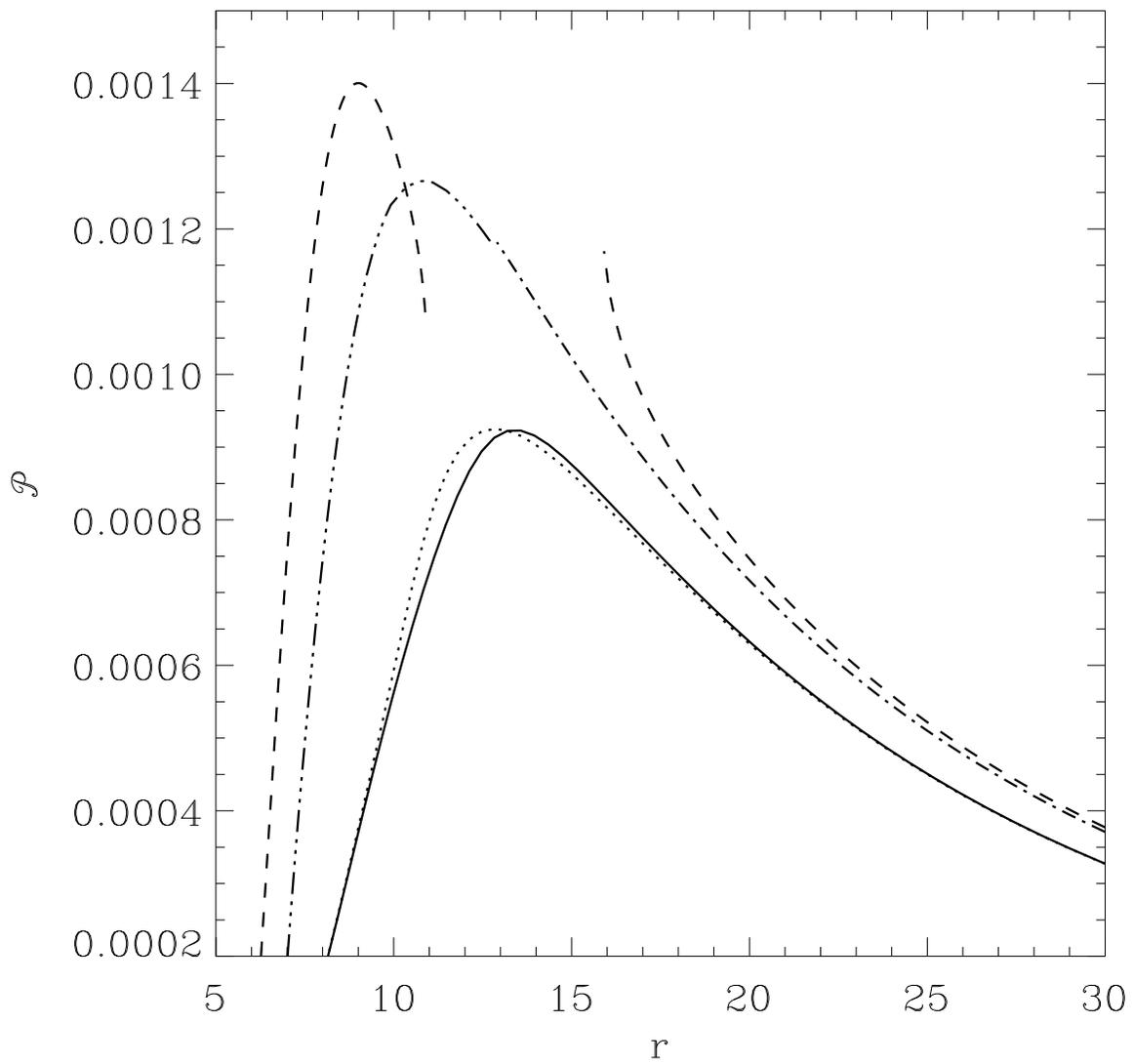
\end{document}